\documentclass[twocolumn]{autart} 
\usepackage{generic}
\usepackage{cite}
\usepackage{amsmath,amssymb,amsfonts}
\usepackage{subfigure}
\usepackage{algorithmic}
\usepackage{empheq}

\usepackage{algorithm}
\usepackage{graphicx}
\usepackage{textcomp}
\usepackage{tikz}
\usetikzlibrary{calc}
\usepackage{cite}
\usepackage{textcomp}
 \usepackage{hyperref}
\usepackage{setspace}
\usepackage{multirow}
\usepackage{xcolor}
\usepackage{bbding}
\usepackage{booktabs}
\usepackage{threeparttable}
\usepackage{bm}
\usepackage{makecell}
\usepackage{tablefootnote}
\def\BibTeX{{\rm B\kern-.05em{\sc i\kern-.025em b}\kern-.08em
 T\kern-.1667em\lower.7ex\hbox{E}\kern-.125emX}}

\usepackage{threeparttable}
\allowdisplaybreaks
\begin{document}
\begin{frontmatter}
\title{Learning Neural Controllers with Optimality and Stability Guarantees Using Input-Output Dissipativity}
\thanks[footnoteinfo]{Keyan Miao and Han Wang contributed equally. Keyan Miao acknowledges support from the Engineering and Physical Sciences Research Council (EPSRC) under project EP/T517811/1, Diego Madeira acknowledges XXXX, Antonis Papachristodoulou acknowledges support from the Engineering and Physical Sciences Research Council grants EP/Y014073/1 and EP/X031470/1.}
\author[eth,oxford]{Han Wang}\ead{hanwang@control.ee.ethz.ch},  % Add the 
\author[oxford]{Keyan Miao}\ead{keyan.miao@eng.ox.ac.uk}, % (ead) as shown
\author[oxford]{Diego Madeira}\ead{diego.madeira@eng.ox.ac.uk},        % e-mail address 
\author[oxford]{Antonis Papachristodoulou}\ead{antonis@eng.ox.ac.uk}        % e-mail address 
\address[eth]{Automatic Control Laboratory (IfA), ETH Z\"{u}rich, Z\"{u}rich, Switzerland}  
\address[oxford]{Department of
Engineering Science, University of Oxford, OX1 3PJ, Oxford, UK}              

\maketitle

\begin{abstract}
Deep learning methods have demonstrated significant potential for addressing complex nonlinear control problems. For real-world safety-critical tasks, however, it is crucial to provide formal stability guarantees for the designed controllers. In this paper, we propose a new framework for designing neural controllers that achieve both stability and optimality with respect to certain functions. Our key idea is to exploit the concept of input-output dissipativity of nonlinear systems by learning neural storage functions and supply rate functions. As a generalization of Lyapunov theory, dissipativity theory provides a natural connection to optimal control theory, offering both stability guarantees and meaningful optimality certificates. The neural controllers can be directly derived from the learned supply rate functions and guarantee closed-loop stability while inheriting optimality properties that can be shaped towards user-defined control objectives. Extensive numerical experiments demonstrate the effectiveness of our approach.
\end{abstract}
\end{frontmatter}

\section{Introduction}
\label{sec:introduction}
% \begin{figure}[ht]
%     \centering
%     % First Subfigure
%     \subfigure[]{
%         \includegraphics[width=0.22\textwidth]{img/level-set.png}
%         \label{fig:first}
%     }
%     \hfill 
%     \subfigure[]{
%         \includegraphics[width=0.22\textwidth]{img/mean-variance-supply.png}
%         \label{fig:second}
%     }
%     \caption{Experimental results for an inverted pendulum. Figure \ref{fig:first} shows the region of attraction (black contour) and the trajectories (colourful lines) of the closed-loop system. All the trajectories converge to zero. Figure \ref{fig:second} presents the evolution of the storage function (in yellow) and the supply rate function (in blue). The value of storage function decreases monotonically to zero, while the supply rate is always negative, representing energy dissipation.}
%     \vspace{-15pt}
%     \label{fig:combined}
% \end{figure}
In recent years, deep learning has become a prevalent method for complex nonlinear control tasks, including robotics \cite{tang2024deep}, intelligent transportation \cite{haydari2020deep}, power systems \cite{zhang2022review}, and others. While these learning-based methods effectively handle complex control tasks that challenge traditional approaches, they in general come with no formal guarantees of desired properties, such as stability.

In control theory, Lyapunov's second method \cite{sastry2013nonlinear} is a widely used method for analyzing stability by designing an energy-like Lyapunov function. Stability is certified if this function strictly decreases along closed-loop trajectories. However, constructing such a function is non-trivial. For polynomial or polynomial-transformable systems, sum-of-squares programming can be used to design stabilizing control laws and Lyapunov functions \cite{yang2023approximate, dai2023convex, anderson2015advances, papachristodoulou2002construction}. Yet, this approach is only immediately applicable to polynomial systems and it is known that some stable polynomial systems may lack a global polynomial Lyapunov function \cite{ahmadi2018globally}.

Learning-based methods for neural controllers and Lyapunov functions is a promising alternative approach \cite{zhou2022neural, chang2019neural, chen2020learning, dai2021lyapunov, yanglyapunov, wu2023neural}. To ensure formal guarantees, these methods employ a counterexample-guided inductive synthesis framework \cite{abate2018counterexample}, with stability verified via tools such as Satisfiable Modulo Theories (SMT) \cite{gao2013dreal}, bound propagation \cite{wang2021beta}, mixed-integer programming \cite{tjeng2017evaluating}, and semi-definite programming \cite{fazlyab2020safety}. While effective for stabilizing control, these approaches may not optimize performance metrics like cumulative energy consumption, as neural controllers are trained over \emph{samples} rather than \emph{trajectories}. This somewhat myopic design lacks infinite-horizon optimality. To bridge this gap, in this paper we develop neural controllers that not only ensure stability but also provide guaranteed optimality over an infinite horizon.

The key technology we are using is dissipativity theory \cite{willems1972dissipative}, which generalizes Lyapunov theory with generalized energy functions, namely storage functions and supply rate functions that depict energy dissipation for input-output pairs. Similarly to Lyapunov's method, dissipativity theory can also be used to certify stability and design stabilizing controllers. A key advantage of dissipativity-based control is its inherent optimality over an infinite horizon \cite{moylan1973nonlinear,grune2022dissipativity}. Recent work has applied this to stochastic nonlinear systems, assuming fixed storage and supply rate functions for Lyapunov-type stabilizing control \cite{lanchares2024nonlinear}. For more flexible and practical design, one can consider learning these functions, though this is even more challenging than Lyapunov function design. The difficulty arises because \emph{dissipativity is an input-output property}, whereas \emph{Lyapunov stability is a state property}. From a learning perspective, this requires sampling over the product of input and output spaces, making the problem especially difficult for multi-input multi-output systems. To address this, we propose a novel reformulation of the dissipativity condition that depends only on the output space, reducing the sampling space's dimensionality.

Previous work on analyzing and learning dissipativity, as well as designing feedback controllers, includes the following. For polynomial systems, sum-of-squares programming has been used to design polynomial storage functions and quadratic supply rate functions \cite{madeira2021necessary,schweidel2021compositional}, making it more applicable to polynomial dynamics, similar to Lyapunov functions. Recently, neural proportional-integral control based on equilibrium-independent passivity was proposed \cite{cui2024structured}, but it assumes known storage and supply rate functions and strictly passive open-loop systems. A similar approach was adopted in \cite{junnarkar2024synthesizing} for dissipativity certification, though it relies on simple quadratic parameterizations. Learning dissipative neural dynamics has been explored in \cite{xu2023learning, li2022learning, okamoto2024learning, tang2024deep}, but these works do not address the more challenging problem of control design. 

Our main contributions are: 
\begin{itemize}
\item We propose a new framework for designing neural control with provable stability guarantees by learning input-output dissipativity. The resulting control law stabilizes \emph{any} system sharing the same dissipativity property;
\item We prove that the designed neural control is the optimal solution to an infinite-horizon optimal control problem, with the cost functional learned based on user-defined objectives;
\item We reformulate the dissipativity conditions to reduce the dimensionality of the sampling space, enabling more efficient learning of neural storage and supply rate functions;
\item We validate the stability and optimality of our method through numerical experiments, demonstrating superior performance over neural Lyapunov-based control in optimality.
\end{itemize}
The paper is organized as follows. Section \ref{sec:problem-statement} formulates the problem of designing optimal and stabilizing control laws and outlines key challenges. Section \ref{sec:methodology} presents our approach to neural control design via input-output dissipativity. The training and formal verification algorithm is detailed in Section \ref{sec:learning-algorithm}, followed by experimental validation in Section \ref{sec:experiments}.
\section{Problem Statement}\label{sec:problem-statement}

Consider a plant with the following nonlinear dynamics
\begin{equation}\label{eq:nominal-system}
    \dot x(t) = f(x(t))+g(x(t))u(t)
\end{equation}
where $x(t)\in\mathcal{X}\subset\mathbb{R}^n$ is the state and $u(t)\in\mathcal{U}\subset\mathbb{R}^m$ is the control input. We assume that the state $x(t)$ of system \eqref{eq:nominal-system} can be measured for any time $t\ge 0$. Functions $f(x):\mathcal{X}\to \mathbb{R}^n$ and $g(x):\mathcal{X}\to \mathbb{R}^{n\times m}$ are assumed to be locally Lipschitz continuous on the compact set $\mathcal{X}$ which has a non-empty interior.
\subsection{Stabilizing Control}
The primary goal of this paper is to design a locally Lipschitz continuous state-feedback controller $\pi(x(t))$ that \emph{stabilizes} the plant \eqref{eq:nominal-system} around the equilibrium point $x^*\in\mathbb{R}^n$. Note that local Lipschitz continuity is essential for local solution uniqueness and existence of the ODE $\dot x =f(x)+g(x)\pi(x(t))$. 
% \begin{defn}[Local Asymptotic Stability]\label{def:las}
%     Consider System \eqref{eq:nominal-system} with a locally Lipschitz continuous controller $u(t)=\pi(x(t))$. The equilibrium point $x^*$ of the system is said to be \emph{locally asymptotically stable} if, $\forall \epsilon>0$ there exists a $\delta>0$ such that if $||x(0)-x^*||<\delta$, then $||x(t)-x^*||<\epsilon$, $\forall t\ge 0$, and $\exists \delta'>0$ such that if $||x(0)-x^*||<\delta'$, then $\lim_{t\to \infty}||x(t)-x^*||=0$. 
% \end{defn}
The key to prove local asymptotic stability of the equilibrium points of nonlinear systems is the existence of local Lyapunov functions.
\begin{defn}[Local Lyapunov Function]\label{def:clf}
    Consider the system \eqref{eq:nominal-system} with a locally Lipschitz continuous controller $u(t)=\pi(x(t))$, and a compact set $\mathcal{X}\subset \mathbb{R}^n$ that contains $x^*$. A function $V(x):\mathcal{X}\to \mathbb{R}$ is called a Lyapunov function for the system if it satisfies
    \begin{align}
        &V(x^*)=0,\quad V(x)>0,\forall x\in\mathcal{X}\backslash \{x^*\} \nonumber\\
        &\dot V(x)=\frac{\partial V(x)}{\partial x}[f(x)+g(x)\pi(x)]<0,\forall x\in\mathcal{X}\backslash\{x^*\}. \label{eq:clf}
        \end{align}
\vspace{-5pt}
\end{defn}
As a consequence of $f(x^*)+g(x^*)\pi(x^*)=0$, we have $\dot V(x)|_{x=x^*}=0$. From the Lyapunov's direct method, the existence of a local Lyapunov function ensures local asymptotic stability of the system \eqref{eq:nominal-system} with control $u(t)=\pi(x(t))$. This motivates the definition of \emph{stabilizing control}. 
\begin{defn}[Stabilizing Control]
    Consider the system \eqref{eq:nominal-system}. The locally Lipschitz continuous control $\pi(x(t))$ is called a \emph{stabilizing control}, if it stabilizes the nonlinear system asymptotically.
    \vspace{-5pt}
\end{defn}
% For stabilizable linear dynamical systems where $f(x)=Ax$ and $g(x)=B$, it is well-known that a quadratic Lyapunov function $x^\top Px$ and a corresponding stabilizing control law $u(t)=Kx(t)$ always exists. The gain $K$ and matrix $P$ can be designed by solving the algebraic Riccati equation that satisfy certain regularity conditions \cite{bertsekas2012dynamic}.
% However, for more general nonlinear systems \eqref{eq:nominal-system}, designing a stabilizing control law and a Lyapunov function to ensure stability is challenging \cite{anderson2015advances}.
\subsection{Optimal Control}
Another problem of interest is optimality. The asymptotically stable optimal control problem is given as follows: 
\begin{align}
    u^*(t)=\mathop{\arg\min}_{u(\cdot)}&\int_0^\infty l(x,u) d t\nonumber\\
    \mathrm{subject~to~}&\dot x = f(x)+g(x)u\label{eq:oc}\\
    &\lim_{t\to \infty}x(t)=0\nonumber
\end{align}
Here, the cost function $l(x,u):\mathbb{R}^{n\times m}\to\mathbb{R}_+$ is the metric for control performance. Take linear quadratic regulator (LQR) problems as an example: the cost function $l(x,u)=x^\top Qx+u^\top Ru$ consists of terms related to state deviations and control effort. The two constraints are clear, one for system evolution and one for stability \cite{sepulchre2012constructive}. We are now at the stage of defining \emph{optimal stabilizing control}.
\begin{defn}[Optimal Stabilizing Control]\label{def:osc}
    Consider System \eqref{eq:nominal-system}. The locally Lipschitz continuous control $\pi(x(t))$ is called an \emph{optimal stabilizing control law} if it is the optimal solution of some optimal control problem \eqref{eq:oc}, i.e. $u^*(t)\equiv\pi(x(t))$.
\vspace{-5pt}
\end{defn}
In control theory, optimal control is widely used for its long-term optimal performance formulation. However, solving \eqref{eq:oc} is arduous for nonlinear systems, even with a quadratic cost functional. Problem \eqref{eq:oc} is an infinite optimization problem, of which the exact solution requires solving the computationally expensive Hamilton-Jacobian-Bellman equation.

% ; 2) The stability constraint $u(\cdot)\in\mathcal{K}(f,g)$ is hard to formulate for nonlinear systems. Even replacing it with a \emph{fixed} Lyapunov function constraint $\frac{\partial V(x)}{\partial x}[f(x)+g(x)u]<0$, solving problem \eqref{eq:oc} is still challenging as the number of inequality constraints is infinite. 
\subsection{Design Goal}
We aim to design a neural control $\pi(x(t))$ as the optimal solution to an infinite-horizon optimal control problem \eqref{eq:oc} for general nonlinear systems, ensuring both optimality and stability by design.
\section{Methodology}\label{sec:methodology}
As discussed in Section \ref{sec:introduction}, previous works learn a neural controller with stability guarantees but do not consider optimality \cite{zhou2022neural, chang2019neural, chen2020learning, dai2021lyapunov, yanglyapunov, wu2023neural}. In this section, we present the theoretic control design framework, then prove stability and optimality of the designed control. Without loss of generality, we consider $x^*=0$, i.e., stabilization around the origin, in the sequel. The results can be easily extended to the case where $x^*\ne 0$.
\subsection{Control with Stability Guarantees}
The key technology we are using is \emph{dissipativity theory}, as a generalization to Lyapunov theory. 
\begin{defn}[Dissipativity]
    Consider system \eqref{eq:nominal-system}. The system is called \emph{dissipative} if there exist functions $V(x):\mathcal{X}\to\mathbb{R}$ and $w(x,u):\mathcal{X}\times\mathcal{U}\to\mathbb{R}$, such that
    \begin{subequations}\label{eq:dissi}
        \begin{align}
            &V(0)=0,\quad V(x)>0,\forall x\in\mathcal{X}\backslash\{0\},\label{eq:dissi-1}\\
            &\frac{\partial V(x)}{\partial x}[f(x)+g(x)u]\le w(x,u),\forall x\in\mathcal{X},u\in\mathcal{U}.\label{eq:dissi-2}
        \end{align}
    \end{subequations}
\end{defn}
$V(x)$ is called a \emph{storage function}, which formulates the totally (nonnegative) energy storage of the system \eqref{eq:nominal-system}, as a generalization of Lyapunov functions. $w(x,u)$ is called \emph{supply rate function}, which represents the supply rate that can be `tolerated' or `dissipated' by the system.

% For the special case where $w(x,0)< 0,\forall  x\in\mathcal{X}\backslash\{0\}$, it is clear that the system is open-loop stable, which means that the system is automatically stable without a control input. Such case has been considered for some power system applications \cite{cui2024structured}, however, it does not hold for general nonlinear systems.\\ 
Dissipativity provides a nice insight for designing stabilizing controllers. Intuitively, if a system is dissipative with a storage function $V(x)$ and a supply rate function $w(x,u)$, the controller $\pi(x)$ should be designed such that
\begin{equation}\label{eq:stable-supplyrate}
    w(x,\pi(x)) < 0, \forall x\in\mathcal{X}\backslash\{0\}.
\end{equation}
Then, the closed-loop system $\dot x =f(x)+g(x)\pi(x)$ is naturally stable as $V(x)$ is a Lyapunov function. The following theorem is the main result of this paper. It proposes a \emph{structure} of the controller $\pi(x)$ that always satisfies \eqref{eq:stable-supplyrate}.
\begin{thm}\label{th:stability}
Consider system \eqref{eq:nominal-system}, a storage function $V(x)$ and a supply rate function
    \begin{equation}\label{eq:qsr}
        w(x,u)=x^\top Q(x)x+2x^\top S(x)u+u^\top R(x)u
    \end{equation}
    where $Q(x):\mathcal{X}
\to\mathbb{R}^{n\times n}, S(x): \mathcal{X}\to \mathbb{R}^{n\times m}, 0\prec R(x):\mathcal{X}\to \mathbb{R}^{n\times n},\forall x\in\mathcal{X}$ that satisfy \eqref{eq:dissi}. If
\begin{equation}\label{eq:delta}
    \Delta(x):=S(x)R(x)^{-1}S(x)^\top - Q(x)\succ 0,\forall x\in\mathcal{X},
\end{equation}
then with
    \begin{equation}\label{eq:structured-control}
        \pi(x)=-R(x)^{-1}S(x)^\top x,
    \end{equation}
the condition \eqref{eq:stable-supplyrate} always holds. $\pi(x)$ is a stabilizing control for the system \eqref{eq:nominal-system}.
\end{thm}
\begin{pf}
    The result provided in the theorem is based on the fact that if (\ref{eq:dissi}) holds with $R(x)\succ0$ and $\Delta (x)\succ 0$, then by considering $u=\pi(x)$, where $\pi(x)$ is given in (\ref{eq:structured-control}), it follows that
\begin{align}\label{eq:sos_50}
\dot{V} (x)\le -x^\top \Delta (x) x < 0, 
\end{align}
which is a closed-loop Lyapunov condition. Thus, the feedback control $\pi(x)$ asymptotically stabilizes the system around the origin. If $V(x)$ is radially unbounded and (\ref{eq:dissi}) is valid $\forall (x,u) \in (\mathbb{R}^n \times \mathbb{R}^m)$, then global asymptotic stability is achieved.

\emph{Necessity} of the conditions mentioned in that theorem can be proved as well, although this is more involved. By assuming that a Lyapunov condition is verified for some $V(x)>0$ and some control law given by 
\begin{align}\label{eq:sos_51}
u(x) = Y(x)x
\end{align}
a solution to dissipation inequality (\ref{eq:dissi}) with $\Delta (x) = 0$ is guaranteed to exist. This means that there is a great degree of generality in the results of Theorem \ref{th:stability}, which justifies its application in this work. 

Firstly, let us suppose that for some control given by (\ref{eq:sos_51}) we have
\begin{align}\label{eq:sos_52}
\nabla V(x)^\top [f(x)+g(x)u(x)]<0.
\end{align}
In addition, assume that there exists some $\bar{M}(x)$ such that
\begin{align}\label{eq:sos_55} 
\nabla V(x) = \bar{M}(x) x,
\end{align}
which can be considered without loss of generality. Then, it also follows that
\begin{align}\label{eq:sos_53}
&\nabla V(x)^\top [f(x)+g(x)\hat{u}(x)] + \\
&\nabla V(x)^\top g(x) \frac{\beta (x)}{4} g(x)^\top \nabla V(x) <0,
\end{align}
with
\begin{align}\label{eq:sos_54}
\hat{u}(x) = K(x)x,~~K(x)=\Big( Y(x)-\frac{\beta(x)}{4} \bar{M}(x) \Big).
\end{align}
\emph{Stability} condition (\ref{eq:sos_53}) leads to
\begin{align}\label{eq:sos_56}
&\nabla V(x)^\top f(x) < \nonumber\\
&-\nabla V(x)^\top g(x) K(x)x-\nabla V(x)^\top g(x) \frac{\beta (x)}{4} g(x)^\top \nabla V(x).
\end{align}

At the same time, the \emph{dissipativity} matrix condition (\ref{eq:matrix-dissi}) can be equivalently presented as
\begin{align}\label{eq:sos_57}
&\nabla V(x)^\top f(x) < \nonumber\\
&x^\top Q(x)x -\nabla V(x)^\top g(x) \frac{R(x)^{-1}}{4} g(x)^\top \nabla V(x) \nonumber\\
&+\nabla V(x)^\top g(x) R(x)^{-1}S(x)^\top x - x^\top S(x) R(x)^{-1}S(x)^\top x.
\end{align}
One can quickly verify that stabilizability implies dissipativity if the right-hand side of condition (\ref{eq:sos_56}) is not grater than the right-hand side of (\ref{eq:sos_57}). This is the case if $R(x) = \beta(x)^{-1} I$, 
\begin{align}\label{eq:sos_58}
S(x)=-\Big[Y(x)-\frac{\beta (x)}{4}M(x) \Big]^\top R(x), \nonumber
\end{align}
$K(x)=-R(x)^{-1}S(x)^\top$ and $Q(x)=S(x)R(x)^{-1}S(x)^\top$, which means that $\Delta(x)=0$.

The proof of Theorem \ref{th:stability} is based on Theorem 1 of \cite{madeira2021necessary}. It provides a powerful insight on the relationship between dissipativity theory and the problem of feedback stabilization. The main idea behind this approach is that a \emph{storage function} may play the role of a \emph{control-Lyapunov function} for the closed-loop system, if certain conditions on the supply rate are satisfied. This provides are great degree of flexibility in designing a controller when compared to passivity-based approaches, for example, where a passive structure must be imposed on the closed loop. In dissipativity-based control, instead, we analyze the open loop properties of the plant; no specific closed-loop behaviour, such as a Hamiltonian structure, needs to be guaranteed in order to obtain stability. \hfill \qed
\end{pf}
In Theorem \ref{th:stability}, the supply rate function $w(x,u)$ is considered in the specific form \eqref{eq:qsr}, generalizing the so-called \emph{QSR supply rate}, where $Q(x)$, $S(x)$, and $R(x)$ are constant matrices. While widely used in convex optimization-based control design \cite{madeira2021necessary}, this formulation can make the dissipativity condition \eqref{eq:dissi} conservative or intractable for general nonlinear systems. In Section \ref{sec:learning-algorithm}, we show how to learn $V(x)$, $Q(x)$, $S(x)$, and $R(x)$ to provably satisfy \eqref{eq:dissi}.

% {\color{yellow}To see this, 
% Then, the dissipativity condition \eqref{eq:dissi-2} always holds for any $R(x)\succ 0$, since $u^\top R(x)u\succ 0,\forall x\in\mathcal{X}$. The existence of such $Q(x)$ and $S(x)$ is guaranteed, as in the following proposition. 
% \begin{proposition}\label{prop:equivalence}
%     Consider the system \eqref{eq:nominal-system} and a storage function $V(x)$ that satisfies \eqref{eq:dissi-1}. Then, there always exists matrices $Q(x)$ and $S(x)$ that satisfy \eqref{eq:equivalence-dissi}.
% \end{proposition}
% \begin{proof}
%     See Appendix.
% \end{proof}
% }
So far we have proved that the structured control $\pi(x)$ in \eqref{eq:structured-control} can stabilize the system. In the next sub-section, we show that it also endows the closed loop system with a notion of optimality, as it is the optimal solution to an optimal control problem with a meaningful cost function.
% \begin{figure*}[h]
%     \centering
%     \input{diagram.tikz}
%     % \vspace{-5pt}
%     \caption{Schematic of the verification-guided dissipativity learning framework for stabilizing and optimal control synthesis}
%     \label{fig:structure}
%     % \vspace{-5pt}
% \end{figure*}
\begin{figure*}[h]
    \centering\includegraphics[width=0.9\textwidth]{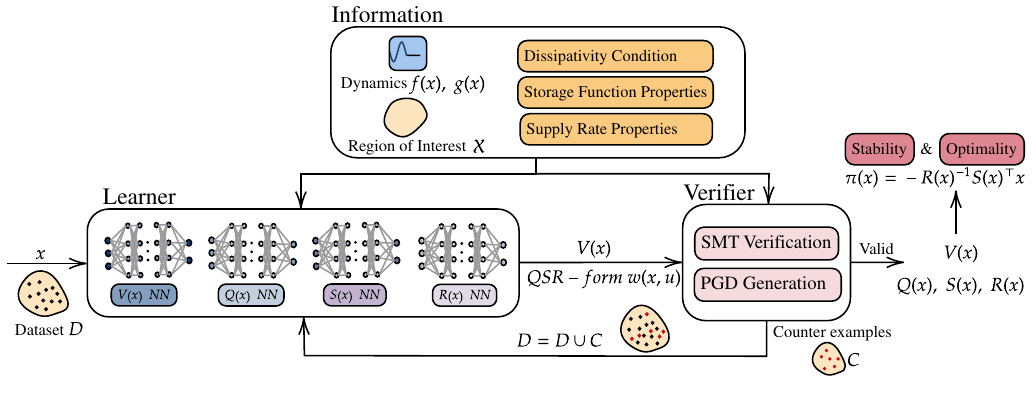}
    % \vspace{-5pt}
    \caption{Schematic of the verification-guided dissipativity learning framework for stabilizing and optimal control synthesis}
    \label{fig:structure}
    % \vspace{-5pt}
\end{figure*}
\subsection{Infinite-time Horizon Optimality Guarantees}
The following theorem demonstrates optimality of $\pi(x)$ as in the form \eqref{eq:structured-control}. 
\begin{thm}\label{th:inverse-optimality}
    Consider the system \eqref{eq:nominal-system}, a storage function $V(x)$ and a supply rate function $w(x,u)=x^\top Q(x)x+2x^\top S(x)u+u^\top R(x)u$, such that \eqref{eq:dissi} and \eqref{eq:delta} hold. Then, $\pi(x)=-R(x)^{-1}S(x)^\top x$ is the optimal solution of the following optimal control problem with positive definite running cost:
    \begin{equation}\label{eq:inverse-oc}
        \begin{split}
\min_{u(\cdot)}&\int_0^\infty \left[\tilde l(x,u)+u^\top R(x)u+x^\top \Delta(x)x\right]dt\\
\mathrm{s.t.~}&\dot x=f(x)+g(x)u\\
&\lim_{t\to \infty}x(t)=0
        \end{split}
    \end{equation}
where $\tilde l(x,u)=-\frac{\partial V(x)}{\partial x}f(x)-\frac{\partial V(x)}{\partial x}g(x)u+x^\top Q(x)x+2x^\top S(x)u$. Moreover, the value function of \eqref{eq:inverse-oc} is $V(x)$.
\end{thm}
\begin{pf}
Substituting
\begin{equation}\label{eq:optimalcontrol}
    v=u+R(x)^{-1}S(x)^\top x 
\end{equation}
into \eqref{eq:inverse-oc} and applying dynamic programming, we have
\begin{align*}
    J&=\min_{u(\cdot)}\int_{0}^\infty [\tilde l(x,u)+x^\top R(x)x+x^\top \Delta(x)x]dt\\
    &=-\min_{u(\cdot)}\int_0^\infty \dot V(x)dt+\min_{v(\cdot)}\int_{0}^\infty [w(x,u)+x^\top\Delta(x)x]dt\\
    &=V(x(0))-\lim_{T\to\infty}V(x(T))+\min_{v(\cdot)}v^\top R(x)v dt
\end{align*}
Given that we are minimizing over these $u$ such that $\lim_{t\to \infty}x(t)=0$, we have $\lim_{T\to\infty}V(x(T))=0$. Moreover, since $R(x)\succ 0$ for any $x$, the minimum of $\min_{v(\cdot)}v^\top R(x)vdt$ is attained at $v=0$. From \eqref{eq:optimalcontrol}, we obtain that the nonlinear control $u=-R(x)^{-1}S(x)^\top x$ is the optimal solution and the storage function $V(x)$ is the optimal value function.

We then show that the cost function of \eqref{eq:inverse-oc} is always positive definite. The cost function can be split into two terms: 1) $\tilde l(x,u)+u^\top R(x)u$ and 2) $x^\top \Delta(x)x$. Term $1)$ is positive semi-definite provided that \eqref{eq:dissi-2} holds, while term $2)$ is positive definite because $\Delta(x)\succ 0$ from \eqref{eq:delta}. \hfill \qed
\end{pf}

The following proposition demonstrates that the cost functional of the optimal control problem becomes nonlinear quadratic if an additional regularity condition is satisfied.
\begin{prop}\label{prop:stronger-optimality}
    Consider the system \eqref{eq:nominal-system}, a storage function $V(x)$ and a supply rate function $w(x,u)=x^\top Q(x)x+2x^\top S(x)u+u^\top R(x)u$, such that \eqref{eq:dissi}, \eqref{eq:delta} and additionally \begin{equation}\label{eq:equivalence-dissi}
    x^\top Q(x)x=\frac{\partial V(x)}{\partial x}f(x),\quad 2x^\top S(x)u=\frac{\partial V(x)}{\partial x}g(x).
\end{equation}
hold. Then, $\pi(x)=-R(x)^{-1}S(x)^\top x$ is the optimal solution of the following optimal control problem:
    \begin{align}\label{eq:inverse-oc-new}
\min_{u(\cdot)}&\int_0^\infty \left[u^\top R(x)u+x^\top \Delta(x)x\right]dt\nonumber\\
\mathrm{s.t.~}&\dot x=f(x)+g(x)u\\
&\lim_{t\to\infty}x(t)=0\nonumber
    \end{align}
Moreover, the value function of \eqref{eq:inverse-oc-new} is $V(x)$.
\end{prop}
The proof of Proposition \ref{prop:stronger-optimality} follows directly from Theorem \ref{th:inverse-optimality}, noting that $\tilde l(x,u)=0$ when \eqref{eq:equivalence-dissi} holds. 

The term $ u^\top R(x) u$ represents the total energy consumption required to stabilize the system from $x(0)$, while $x^\top \Delta(x)x$ penalizes deviations from the origin. The cost function can be adjusted by tuning $R(x) \succ 0$, $\Delta(x)$, and $V(x)$. A larger $R(x)$ prioritizes control effort minimization, leading to smoother but slower convergence. Conversely, a smaller $R(x)$ emphasizes state deviation minimization, resulting in faster convergence but higher energy consumption.
% \begin{figure*}
%     \centering
%     \includegraphics[width=1\linewidth]{img/diagram-20250122.pdf}
%     \caption{\textcolor{blue}{TBD-adjust the tikz - too long, font bold, layout}
%     {\color{red} Is it possible to show optimality as well? For example, draw an optimal control problem on top of the figure. Although the storage function is parameterized by $V(x)=x^\top \tilde V(x)x$ where $\tilde V(x)$ is a NN, I think we can still use $V(x)$ instead of $\tilde V(x)$ in the figure to avoid misunderstanding.}}
%     \label{fig:structure}
% \end{figure*}
% \begin{remark}\label{rem:optimality}
%     Existing literature \cite{zhou2022neural, chang2019neural, chen2020learning, dai2021lyapunov, yanglyapunov, wu2023neural} \footnote{Some of these works consider discrete-time dynamics, but the optimality analysis remains similar to the continuous ones.} on learning a neural stabilizing control $\zeta(x)$ together with a neural Lyapunov function has no proved inverse optimality guarantees. From the control perspective, this is because the output feedback dynamics
% \begin{align*}
%     &\dot x = f(x)+g(x)u,\\
%     &y=\zeta(x)
% \end{align*}
% do not have output feedback passivity guarantees, according to Definition 2.12 and Theorem 3.2 of \cite{sepulchre2012constructive}. We give more explanations in Appendix \ref{sec:optimality}.
% \end{remark}
\section{Counterexamples Inductive Learning Algorithm}\label{sec:learning-algorithm}
% \textcolor{blue}{The structure of this section:
% \begin{itemize}
%     \item Introduce the whole framework including learning and verification
%     \item The training procedure (mainly the design of loss function)
%     \item The verification process including SMT and PGD
% \end{itemize}}
In this section, we present the main algorithm of designing the storage function $V(x)$ and the supply rate function $w(x,u)=x^\top Q(x)x+2x^\top S(x)u+u^\top R(x)u$ that provably satisfy the conditions:
\begin{itemize}
    \item Storage function properties:\begin{itemize}
        \item $V(0)=0$
        \item The bi-Lipschitz condition: $\mu x^\top x \le V(x)\le \nu x^\top x,\forall x\in\mathcal{X}$
    \end{itemize}
    \item QSR-form supply rate properties:\begin{itemize}
        \item $R(x)\succ 0$
        \item $\Delta(x):=S(x)R(x)^{-1}S(x)^\top - Q(x)\succ 0$
    \end{itemize}
    \item The dissipativity condition:
    $$\frac{\partial V(x)}{\partial x}[f(x)+g(x)u]\le w(x,u),\forall x\in\mathcal{X},u\in\mathcal{U}.$$
\end{itemize}
and additionally
\begin{itemize}
    \item Cost functional shaping w.r.t. a user defined objective $l(x,u)$:
    \begin{equation*}
        \tilde l(x,u)+u^\top R(x)u+x^\top \Delta(x)x \approx l(x,u),\forall x\in\mathcal{X},\forall u\in\mathcal{U}
    \end{equation*}
\end{itemize}
Figure \ref{fig:structure} shows the learning structure. The main idea is to use the counter example guided inductive synthesis framework \cite{abate2018counterexample}. Our framework consists of two main components working together iteratively:
\begin{itemize}
    \item A learning module that trains neural networks to represent $V(x)$, $Q(x)$, $S(x)$, and $R(x)$, optimizing over data samples to satisfy the required conditions. The learning is guided by carefully designed loss functions.
    \item A verification module that employs  Satisfiability Modulo Theory (SMT) solvers to formally verify whether the learned functions satisfy all requirements. When violations are found, it generates counterexamples for improving the learned functions.
\end{itemize}
When the verifier finds a violation of any condition, the counterexample point is used by the learner to update the neural networks, focusing the training on problematic regions of the state space. This process continues until all conditions are provably satisfied. In the following subsections, we present our approach for learning storage and supply rate functions, and the complete verification procedure.
\subsection{Learner of Storage and Supply Rate Functions}
% \textcolor{blue}{To Do: make it more clear that what are the conditions we need to satisfy, may use bullet points? 1) On the properties of storage $V(x)$; 2) On the properties of $Q$ $S$ and $R$; 3) On the dissipativity condition - I put these before this subsection, but can also be moved here}
\textbf{Dissipativity condition:} Similarly to existing work on learning neural Lyapunov functions, $V(x)$, $Q(x)$, $S(x)$, and $R(x)$ are parameterized by feed-forward neural networks. However, unlike Lyapunov function training, which relies on \emph{state samples} to satisfy stability conditions \eqref{eq:clf}, the dissipativity condition \eqref{eq:dissi-2} requires training over \emph{state-input samples}. This poses sampling challenges for systems with high control dimension $m$ (e.g., fully or overly actuated systems). To mitigate this, we introduce a novel reformulation of \eqref{eq:dissi-2} that reduces the sampling space dimension.
\begin{prop}
    Consider the system \eqref{eq:nominal-system}, a storage function $V(x)$ and a supply rate function $w(x,u)=x^\top Q(x)x+2x^\top S(x)u+u^\top R(x)u$. If for all $x\in\mathcal{X}$, it holds that
    \begin{equation}\label{eq:matrix-dissi}
        \underbrace{\begin{bmatrix}
            R(x)&S(x)^\top x-0.5g(x)^\top \frac{\partial V(x)}{\partial x}^\top\\
            \star & -\frac{\partial V(x)}{\partial x}f(x)+x^\top Q(x)x
        \end{bmatrix}}_{:=M(x)}\succ 0.
    \end{equation}
Then, \eqref{eq:dissi-2} holds.
\end{prop}
\begin{pf}
    By multiplying $[u^\top\quad 1]$ and $[u^\top\quad 1]^\top$ on the left hand side and right hand side of the matrix $M(x)$ in \eqref{eq:matrix-dissi}, we obtain the dissipativity condition \eqref{eq:dissi-2}.
\end{pf}
Although \eqref{eq:matrix-dissi} is a matrix positive definite condition, it only depends on $x$, therefore the dimension of sampling space has been reduced. We also note that the practical meaning of this transformation goes beyond computational efficiency. Consider the original dissipativity condition \eqref{eq:dissi-2}; the sampling space for the states can be defined by the practitioners. However, one may lose the freedom of choosing the sampling space for the input. This is because the magnitude of the structured control $\pi(x)$ is not known \textit{a priori}, but instead \textit{a posteriori}. If the input sampling space is too small, $\pi(x)$ can lie out of this space, while if the sampling space is too large, redundant samples will be unavoidably added. Our formulation addresses this directly.

The loss functional $\mathcal{L}_{d}(\theta;\mathcal{D})$ for the condition \eqref{eq:matrix-dissi} is defined as
\begin{equation}\label{eq:loss-dissipativity}
\mathcal{L}_d(\theta;\mathcal{D}):=\frac{1}{N}\sum_{x_i\in\mathcal{D}}\mathrm{ReLU}(\max\mathrm{eig}(-M(x_i))),
\end{equation}
where $\max\mathrm{eig}(\cdot)$ is the operator for the maximal eigenvalue, $\theta$ is the parameter for the trainable neural networks, $\mathcal{D}$ is the set of state samples and $N$ is the number of samples. When $M(x_i)\succ 0,\forall x_i\in\mathcal{D}$, the maximal eigenvalue for each $-M(x_i)$ is negative, then $\mathcal{L}_d=0$. Otherwise, we have $\mathcal{L}_d>0$. Note that the logarithm determinant barrier function $-\log\det(\cdot)$ has been widely-used in convex optimization literature for positive semi-definite constraints \cite{boyd2004convex}. The $\max\mathrm{eig}(\cdot)$ is also convex, and always well-defined for any real symmetric matrix.

% \vspace{-10pt}
\textbf{Storage function properties:} We then consider the condition \eqref{eq:dissi-1}, which requires the storage function $V(x)$ to be locally positive definite. $V(x)$ is parameterized by 
\begin{equation}\label{eq:storage-func}
    V(x)=x^\top \tilde V(x) x,
\end{equation}
where $\tilde V(x):\mathcal{X}\to\mathbb{R}^{n\times n}$ is parameterized by a feed-forward neural network. Under this parameterization, we immediately have that $V(0)=0$ holds. Instead of enforcing $V(x)$ to be positive-definite, we consider a stronger condition which requires $V(x)$ to be bi-Lipshitz:
\begin{equation*}
    \mu x^\top x \le V(x)\le \nu x^\top x
\end{equation*}
where $0<\nu < \mu$ are constants. The role of the lower bound $\mu x^\top x$ guarantees $V(x)$ is \emph{strictly positive} for $x\ne 0$. Moreover, it eliminates local minima due to numerical precisions, in the sense that the value of $V(x)$ will not be too small for a state that is far away from the origin. On the contrary, the upper bound $\nu x^\top x$ guarantees that the value of $V(x)$ will not be too large for small $x$.\\
This loss functional $\mathcal{L}_v$ for the condition \eqref{eq:dissi-1} is given by
\begin{align}\label{eq:loss-storage}
    \mathcal{L}_v(\theta;\mathcal{D}):=&\frac{1}{N}\sum_{x_i\in\mathcal{D}}\left\{\mathrm{ReLU}(\mu x_i^\top x_i-V(x_i))\right.\nonumber\\
    &\left.+\mathrm{ReLU}(V(x_i)-\nu x_i^\top x_i)\right\}.
\end{align}
% Here, we use two ReLU functions to incorporate the lower and upper bounds for the storage function $V(x)$.
\vspace{-10pt}

\textbf{QSR-form supply rate properties:} The next one is $\Delta(x)\succ 0$ in \eqref{eq:delta}. Similarly to the treatment of the matrix inequality \eqref{eq:matrix-dissi}, we introduce the following loss functional
\begin{equation}
    \mathcal{L}_t(\theta;\mathcal{D}):=\frac{1}{N}\sum_{x_i\in\mathcal{D}}\mathrm{ReLU}(\max\mathrm{eig}(-\Delta(x_i))).
\end{equation}
The last constraint is $R(x)\succ 0$, which is integrated by the following loss functional
\begin{equation}
    \mathcal{L}_r(\theta,\mathcal{D}):=\frac{1}{N}\sum_{x_i\in\mathcal{D}}\mathrm{ReLU}(\max\mathrm{eig}(-R(x_i))).
\end{equation}
The overall loss function for learning the storage function and supply rate function is given by
\begin{equation}\label{eq:loss-func}
    \mathcal{L}(\theta,\mathcal{D}):=w_1\mathcal{L}_d+w_2\mathcal{L}_v+w_3\mathcal{L}_t+w_4\mathcal{L}_r,
\end{equation}
where $w_1,\ldots,w_4>0$ are weights. 
% From the experiments, it is enough to set balanced weights, e.g. $w_1=w_2=w_3=w_4$.
% \vspace{-5pt}

\textbf{Cost functional shaping}One additional objective to be considered is cost functional shaping. In practice, one may want to optimize over a certain cost function $l(x,u)$ rather than the one derived in \eqref{eq:inverse-oc}. Let $\bar l(x,u):=\tilde l(x,u)+u^\top R(x)u+x^\top \Delta(x)x$ denote the cost function of \eqref{eq:inverse-oc}. Then, the cost function shaping can be achieved by considering a new loss function term with a set of sampled input-state $\mathcal{D}'$:
\begin{equation}\label{eq:shaping}
    \mathcal{L}_c=\frac{1}{N}\sum_{x_i,u_i\in\mathcal{D'}}\left( l(x_i,u_i)-\bar l(x_i,u_i)\right)^2.
\end{equation}
\\
This additional loss can be integrated into $\mathcal{L}$ with a weight $w_5$. When $l(x,u)$ is quadratic in $u$, the input samples $u_i$ is not needed for $\mathcal{L}_c$, as one can directly minimize the discrepancy for the coefficients of $u$ in $l(x,u)$ and $\bar l(x,u)$. 
\subsection{Complete Verification and Counterexample Generation}
\begin{algorithm}[tb]
   \caption{Training Structured Neural Controllers}
   \label{alg:example}
\begin{algorithmic}[1]
   \STATE {\bfseries Input:} training data set $\mathcal{D}$, {region of interest $\mathcal X$}, plant dynamics $f(x)$ and $g(x)$, PGD stepsize $\alpha$ and learning rate $\gamma$.
   \STATE {\bfseries Output:} controller $\pi(x)$, storage function $V(x)$, QSR-form supply rate function $w(x,u)$
   \STATE {\bfseries Initialize:} Neural networks $Q(x),S(x), R(x),\tilde V(x)$ {parameterized with $\theta$}
   \REPEAT
   \FOR{$iter=1,2,\ldots$}
   \STATE Randomly sample states $\tilde{x}_j\in\mathcal{X}$
   \FOR{$j=1,2,\ldots$}
      \STATE $\tilde{x}_j\leftarrow \mathrm{Project}_{\mathcal{X}}\left(\tilde{x}_j+\alpha \cdot\frac{\partial \mathcal{L}(\theta;\tilde{x}_j)}{\partial \tilde{x}}\right)$ 
    \ENDFOR
    \STATE $\mathcal{D}\leftarrow \{\tilde x_i\}\cup\mathcal{D}$
   \FOR{$epoch=1,2,\ldots$}
   \STATE $\theta\leftarrow \theta-\gamma \nabla_\theta \mathcal{L}(\theta,\mathcal{D})$
   \ENDFOR
   \ENDFOR
   \UNTIL{Passed the verifier}
\STATE {$\pi(x)\leftarrow -R(x)^{-1}S(x)^\top x$}
\end{algorithmic}
\end{algorithm}
% {\color{red}Instead of discussing dreal, should we also say other verifiers such as crown can be applied to our problem as well? One way could be we do not say we emply dreal for our problem, just say any verifier. The PGD attack is then natural as: 1) dReal is inefficient; 2) crown can not produce counterexamples. We will say we are using dReal in the experiment part. If you felt this is a bad way, I am happy to use write code for crown, this will not be so difficult.}
We employ SMT (Satisfiability Modulo Theory) solvers to verify that our learned storage function $V(x)$ and QSR-form supply rate function $w(x,u)$ are valid and satisfy the dissipativity conditions. Using SMT solvers is a powerful technique for checking whether mathematical formulas hold over the domain of interest.
% These solvers can either prove that conditions are satisfied for all points in the domain or provide counterexamples where the conditions are violated. 
For each candidate storage function $V(x)$ and supply rate $w(x,u)$, we verify three key conditions: 1) the bi-Lipschitz condition for $V(x)$; 2) the dissipativity condition $\dot V(x,u)\le w(x,u)$; 3) the supply rate condition $R(x)\succ 0,\Delta(x)\succ 0$.

The existing SMT solvers such as dReal \cite{gao2013dreal} do not directly support semi-definite constraints in the form of \eqref{eq:matrix-dissi}. To overcome this problem, we reformulate the matrix inequality constraints \eqref{eq:matrix-dissi}, \eqref{eq:delta} and $R(x)\succ 0,\forall x\in\mathcal{X}$ into equivalent and verifiable algebraic constraints. 

For \eqref{eq:matrix-dissi}, consider the Schur complement for the first block $R(x)$, we have an equivalent but verifiable condition
\begin{equation}
\begin{split}
    &-\frac{\partial V(x)}{\partial x}f(x)+x^\top Q(x)x-S(x)^\top x\\
    &+0.5g(x)^\top \frac{\partial V(x)}{\partial x}^\top R(x)^{-1}\star>0,\forall x\in\mathcal{X}
\end{split}
\end{equation}
The constraint \eqref{eq:delta} can be transformed into
\begin{equation}
    x^\top \Delta(x)x> 0,\forall x\in\mathcal{X}.
\end{equation}
The constraint $R(x)\succ 0,\forall x\in\mathcal{X}$ can be transformed into 
\begin{equation}\label{eq:r-verification}
    x^\top R(x)x>0,\forall x\in\mathcal{X}.
\end{equation}
When $R(x)$ is taken as a constant matrix instead of a functional one, the positive definiteness can directly be checked by computing the eigenvalues.

For other verifiers such as mixed-integer programming \cite{tjeng2017evaluating} and $\alpha\beta$-crown \cite{wang2021beta}, the loss function $\mathcal{L}$ in \eqref{eq:loss-func} can directly be used for verification.

While SMT solvers can provide formal guarantees and counterexamples, the verification process can be computationally intensive due to complex symbolic calculations. Some verifiers like $\alpha \beta$-crown offer efficient verification but may not support counterexample generation. Therefore, similar to existing works \cite{yanglyapunov, wu2023neural}, we employ a hybrid approach: before invoking computationally expensive SMT verification, we first use projected gradient descent (PGD) \cite{madry2017towards} to efficiently search for potential counterexamples by maximizing the violation of our conditions. For a given point $\tilde x_j$, we compute the gradient of our conditions $\mathcal{L}(\theta;\tilde x_j)$ with respect to $\tilde x_j$ and take steps in the direction that potentially increases violation, while projecting back onto our domain of interest after each step.

The PGD-based search serves as a fast preliminary check before formal verification. When PGD finds points that likely violate our conditions, these points are added to the training set. This hybrid strategy significantly reduces the computational burden while maintaining the soundness of our approach: PGD helps quickly identify problematic regions for learning, while formal verification is used periodically to provide guarantees. The complete verification-guided learning process alternates between: 1) Using PGD to efficiently find potential violations; 2) Updating the neural networks based on these violations; 3) Periodically employing formal verification for the learned functions.

The training algorithm is shown in Algorithm \ref{alg:example}. On Lines 6-10, counterexamples are generated using PGD and added to the training dataset $\mathcal{D}$. On Lines 11-13, the parameters of neural networks are updated. If the trained neural networks pass verification, the control law $\pi(x)=-R(x)^{-1}S(x)^\top x$ is guaranteed to stabilize the system.
\section{Experiments}\label{sec:experiments}
We evaluate the performance of our proposed framework across three benchmark environments: \textit{Electric Circuit}, \textit{Inverted Pendulum}, and \textit{Rod on a Cart}. Each experiment is designed to assess the framework’s stability guarantees and optimality in comparison with baseline approaches. The controller is trained using Algorithm \ref{alg:example}, and the cost functional shaping is defined as in \eqref{eq:shaping}. Source code is available at \href{https://anonymous.4open.science/r/Learning-Dissipativity-Control-07B6}{https://anonymous.4open.science/r/Learning-Dissipativity-Control-07B6}. 

We compare our method against:\begin{itemize}
    \item Neural Control Lyapunov Function (NCLF) \cite{wu2023neural,yanglyapunov,chang2019neural} : While these methods differ in their choice of verifiers and conditions for enlarging the region of attraction, they share the same fundamental approach of jointly learning  both the controller and Lyapunov function to guarantee stability;
    \item NODE: Optimal control via Neural ODEs;
    \item OURS-NCFS: Our proposed dissipativity-based approach without the cost functional shaping term \eqref{eq:shaping}.
\end{itemize}
The controllers designed by OURS, NCLF, and OURS-NCFS are verified using dReal \cite{gao2013dreal} under the same precision and region. All methods share identical training settings (e.g., optimizer, PGD step size, learning rate, and network sizes).\\
Method NODE designs the control by solving the optimal control problem
\begin{align*}
\pi(x)=\mathop{\arg\min}_{u(\cdot)}~&\int_0^\infty l(x,u)dt,\\
        \mathrm{subject~to}~&\dot x = f(x)+g(x)u,
\end{align*}
where $l(x,u)$ is the user-defined objective function. Our method considers the shape of the loss function $\bar l(x,u)$ with respect to $l(x,u)$ via the loss function term \eqref{eq:shaping}. In the following experiments, the performance oriented cost function $l(x,u)$ is designed as
\begin{equation}\label{eq:cost}
    l(x,u):=x^\top x+u^\top u
\end{equation}

\begin{table*}[t]
\centering
\begin{tabular}{ccccccc}
\hline
Plant    & Controller & $V(x)$ & $Q(x)$ & $S(x)$ & $R(x)$ & Region of interest \\ \hline
Pendulum & Static     & 8      & 8      & 8      & Linear & $[\pi, 2]$         \\
Pendulum & Dynamic    & 32     & 32     & 32     & Linear & $[\pi, 2, 1]$      \\
RodCart  & Static     & 20      & 20      & 20      & Linear & $[0.1, 0.1, 0.1, 0.1]$     \\
Circuit  & Static     & 8      & 8      & 8      & Linear & $[1, 1]$           \\ \hline
\end{tabular}
\caption{Size of neural networks, type of controller and the region of interest for each task.}
\label{tab:nns}
\end{table*}

The size of neural networks and region of interest for each task is shown in Table \ref{tab:nns}. By the number for neural networks $V(x)$, $Q(x)$, $S(x)$ we mean the number of neurons. All of these functions are parameterized by feed-forward neural networks with one hidden layer, using $\tanh$ as the activation function. For all the experiments, the matrix $R(x)$ are parameterized by one linear layer. 

For method ``OURS", the loss function is defined by 
\begin{equation*}
    \mathcal{L}(\theta,\mathcal{D})=w_1\mathcal{L}_d+w_2\mathcal{L}_v+w_3\mathcal{L}_t+w_4\mathcal{L}_r+w_5\mathcal{L}_c
\end{equation*}
where the loss functions are given by \eqref{eq:loss-storage} - \eqref{eq:shaping}. The neural control is then trained using Algorithm \ref{alg:example}. For method ``OURS-NCFS", the loss function is defined by
\begin{equation*}
    \mathcal{L}(\theta,\mathcal{D})=w_1\mathcal{L}_d+w_2\mathcal{L}_v+w_3\mathcal{L}_t+w_4\mathcal{L}_r,
\end{equation*}
and the neural control is trained using Algorithm \ref{alg:example} with the same hyperparameters as ``OURS". For method ``NCLF", the neural Lyapunov function is parameterized in the same way as the storage function for ``OURS" and ``OURS-NCFS". The neural controls for methods ``NCLF" and ``NODE" are trained as neural networks that always have the same numbers of hidden layers and neurons as $S(x)$. This guarantees that the neural controls for all the methods have the same numbers of parameters.

dReal \cite{gao2013dreal} is selected as the verifier in Algorithm \ref{alg:example}. The numerical precision $\delta$ is set to be $10^{-3}$ for all the experiments. We want to highlight here our method \emph{does not} depend on the specific verifier, allowing the users to implement their own verification methods for formal dissipativity and stability guarantees. 

To demonstrate optimality, we consider the following user-defined objective function 
\begin{equation}\label{eq:control-objective}
    \int_0^\infty (l(x,u):=x^\top x+u^\top u)dt
\end{equation}
where $x=[\theta\quad \dot \theta]$ is the state, and $u$ is the control input. After training the four controls, we evaluate the objective function \eqref{eq:control-objective} along the four closed-loop systems:
\begin{equation}\label{eq:odes}
    \dot x = f(x)+g(x)\pi_{\{\cdot\}}(x),
\end{equation}
where $\{\cdot\}\in\{\mathrm{OURS,NCLF,OURS-NCFS,NODE}\}$. In practice, the ordinary differential equations \eqref{eq:odes} are digitalized and solved using fourth-order Runge-Kutta (rk4) method. For each system, the temporal state-input pairs are sampled as
\begin{equation}\label{eq:sequence}
    \{x(0),\pi_{\{\cdot\}}(0)\},\{x(1),\pi_{\{\cdot\}}(1)\},\ldots,\{x(T),\pi_{\{\cdot\}}(T)\},
\end{equation}
and the empirical value is evaluated by
\begin{equation}\label{eq:empirical}
    \sum_{t=0}^T x(t)^\top x(t)+\pi_{\{\cdot\}}(t)^\top \pi_{\{\cdot\}}(t).
\end{equation}
The sequence \eqref{eq:sequence} is sampled to be long enough $T >>0$, such that $x(T)=0,\pi_{\{\cdot\}}(T)$ is ensured. 
% \usepackage{multirow}
\subsection{Electrical Circuit}

\begin{figure}[ht]
    \centering
    \subfigure[Electrical circuit]{
        \includegraphics[width=0.22\textwidth]{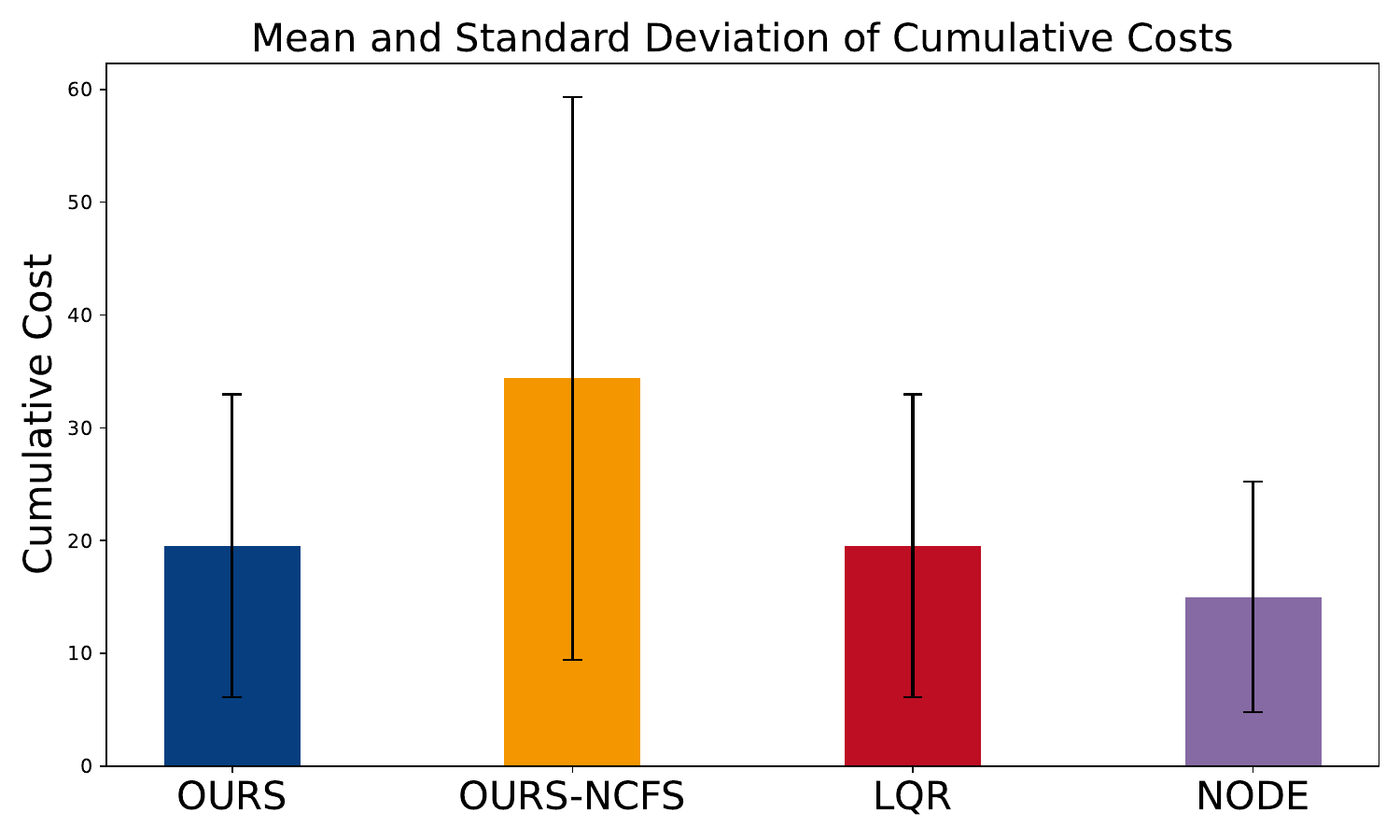}
        \label{fig:circuit-cost}
    }
    \hfill
    \subfigure[Inverted pendulum]{
        \includegraphics[width=0.22\textwidth]{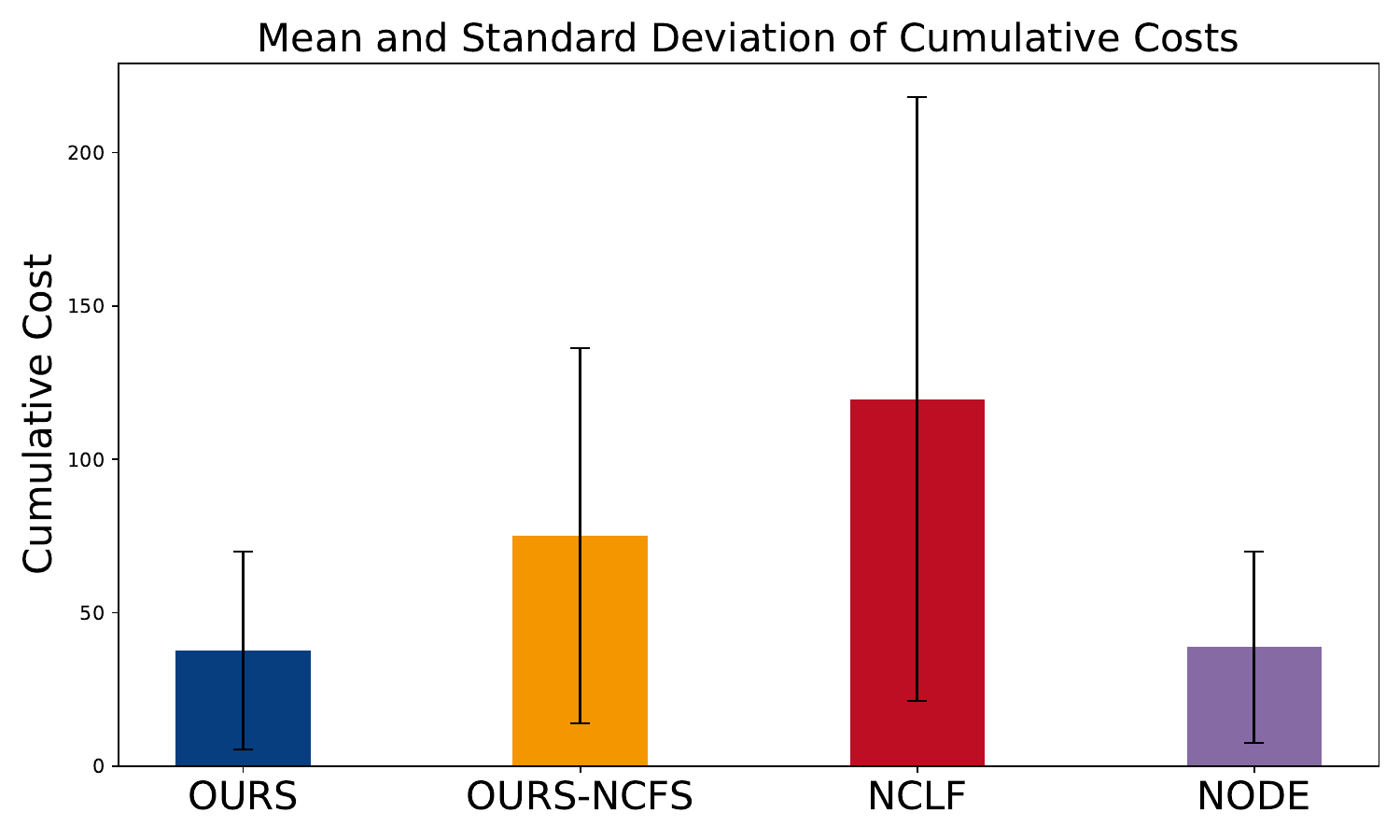}
        \label{fig:ip-cost}
    }
    \caption{The average empirical cumulative cost with standard deviation on 30 trials starting from random initial states.}
    \label{fig:cost-main}
\end{figure}

In the electrical circuit experiment, we evaluate the stability and dissipativity properties of the control designed by our method. Since this is a linear system, we compare our method to the Linear Quadratic Regulator (LQR) with cost function \eqref{eq:cost}, the theoretical optimal control solution for such systems. By applying our cost shaping method, we find that our framework successfully replicates the LQR solution as shown in Figure \ref{fig:circuit-cost} and Table \ref{table:cost}, achieving not only stability but also optimality with respect to the specified performance metric. While NODE appears to achieve a lower cost when evaluated on specific trajectory lengths during training and testing, the learned controller \emph{fails to stabilize the system}. This exposes a key limitation of trajectory-based training: it optimizes performance over predefined trajectories but lacks guarantees of stability for the entire state space. In contrast, our dissipativity-based framework explicitly enforces stability guarantees, ensuring that the system converges robustly, even beyond the training scenarios.

\begin{figure}[ht]
    \centering
    \subfigure[On the $\theta-\dot \theta$ plane]{
        \includegraphics[width=0.22\textwidth]{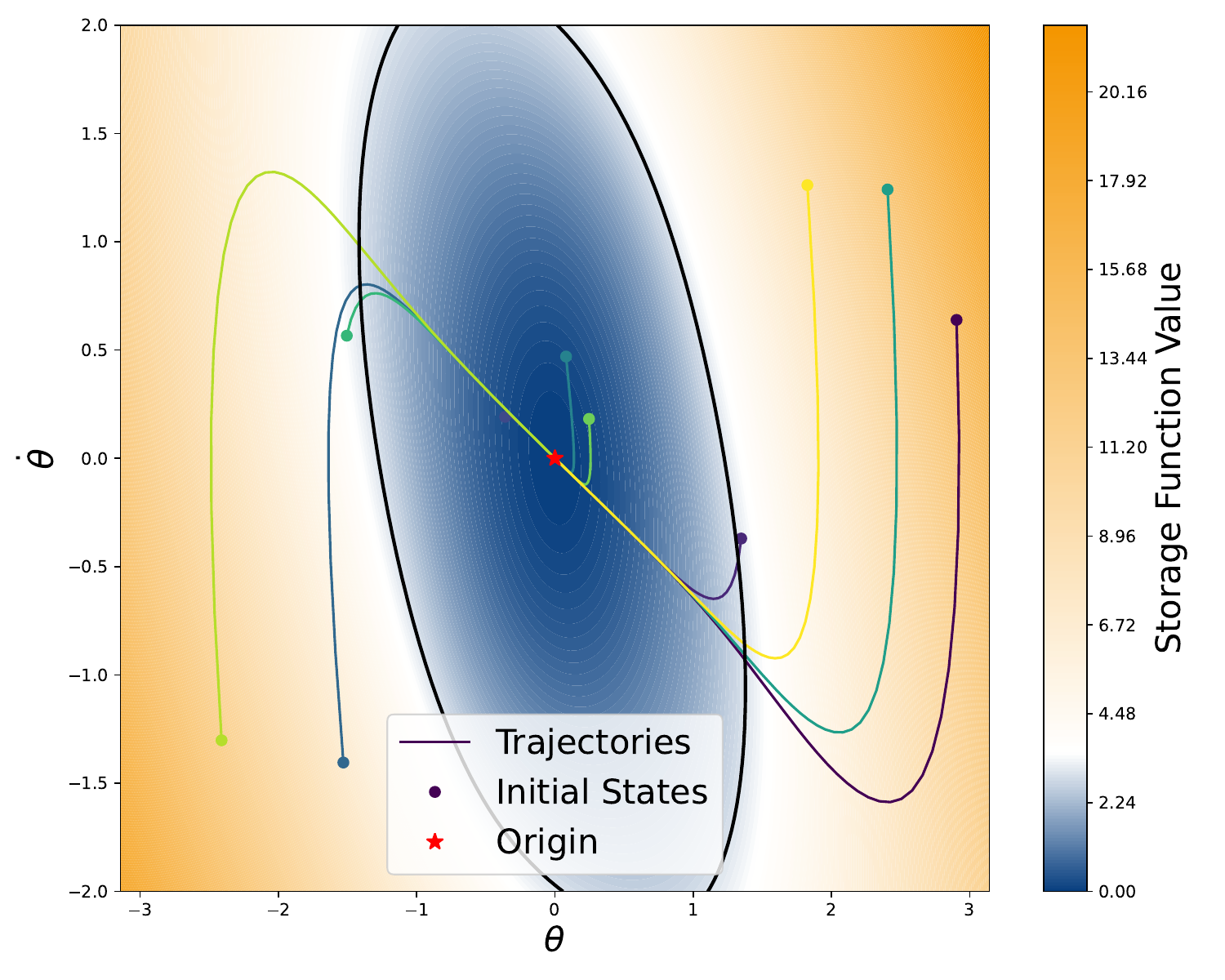}
        \label{fig:IPd-xy}
    }
    \hfill
    \subfigure[On the $\theta-x_c$ plane]{
        \includegraphics[width=0.22\textwidth]{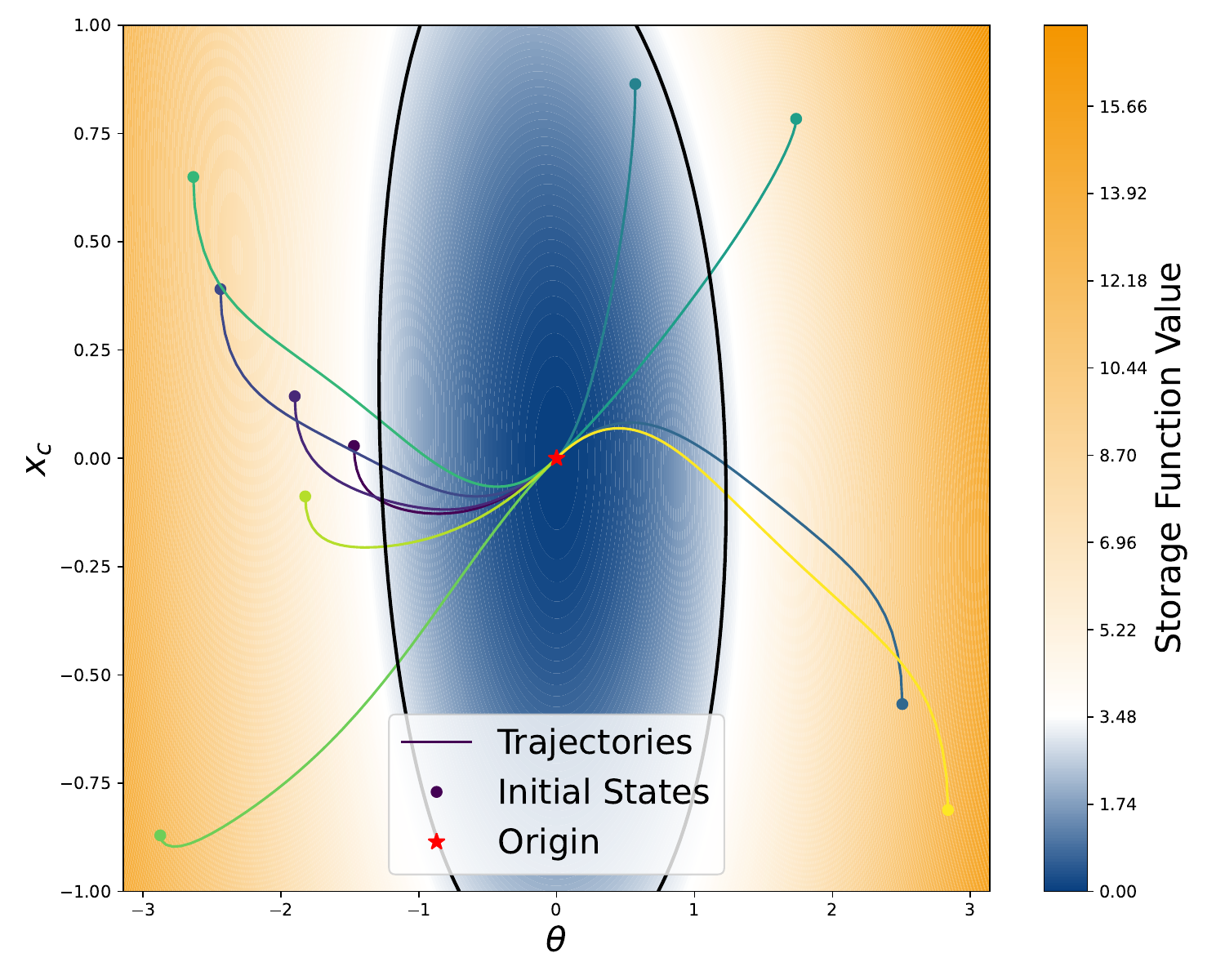}
        \label{fig:IPd-xz}
    }
    \caption{The region of attraction (black contour) and the trajectories (colourful lines) of the closed-loop system.}
    \label{fig:IPd-storage}
\end{figure}
\vspace{-5pt}
\begin{table*}[]
\caption{Empirical cumulative costs across different environments, averaged over 30 trials per method}
\centering
\begin{tabular}{c|cccc|c}
\hline
\multirow{2}{*}{}                & \multicolumn{4}{c|}{Cost}                                                                                                     & \multirow{2}{*}{Stability Guarantee} \\ \cline{2-5}
                & \multicolumn{1}{c|}{Circuit} & \multicolumn{1}{c|}{Pendulum-static} & \multicolumn{1}{c|}{Pendulum-dynamic} & Rod on the Cart &                                      \\ \hline
NCLF            & \multicolumn{1}{c|}{19.5 (LQR)}        & \multicolumn{1}{c|}{119.7}                & \multicolumn{1}{c|}{125.7}                 &           3191.2      & \Checkmark                         \\ \hline
Ours-NCFS & \multicolumn{1}{c|}{34.4}        & \multicolumn{1}{c|}{75.2}                & \multicolumn{1}{c|}{42.5}                 &           412.9      & \Checkmark                         \\ \hline
Ours    & \multicolumn{1}{c|}{19.5}        & \multicolumn{1}{c|}{\textbf{37.8}}                & \multicolumn{1}{c|}{15.1}                 &         361.3        & \Checkmark                         \\ \hline\hline
NODE            & \multicolumn{1}{c|}{\textbf{15.0}}        & \multicolumn{1}{c|}{38.8}                & \multicolumn{1}{c|}{\textbf{13.7}}                 &          \textbf{284.0}       & \XSolid               \\ \hline
\end{tabular}
\label{table:cost}
\end{table*}
\begin{figure*}[ht]
    \centering
    \includegraphics[width=0.9\linewidth]{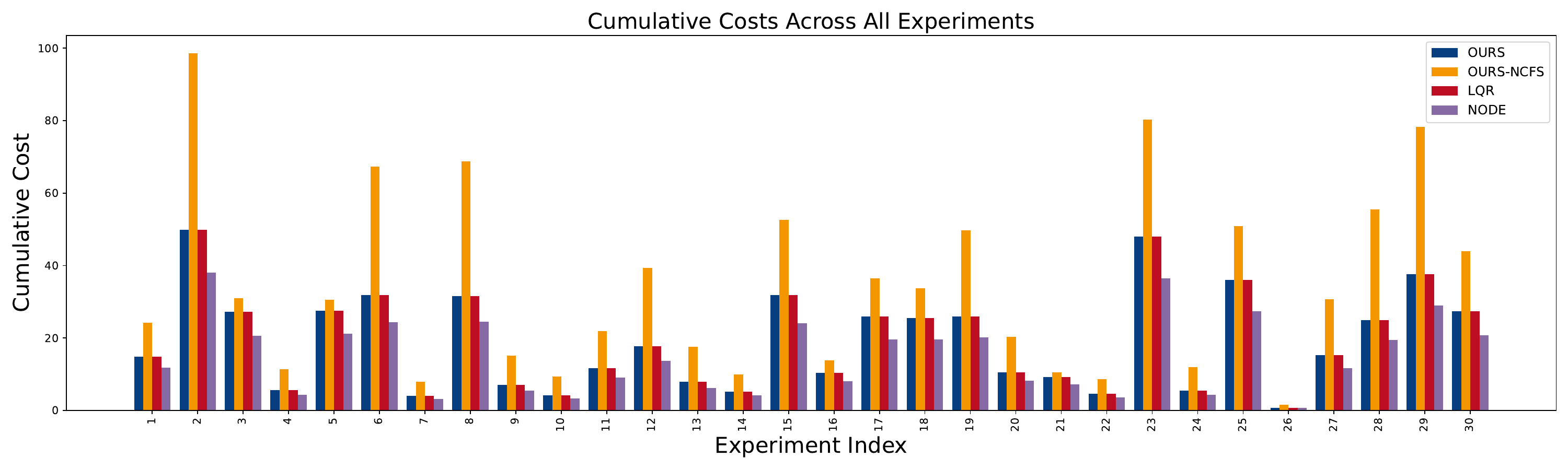}
    \caption{Comparison of empirical cumulative costs \eqref{eq:empirical} for the Electrical Circuit with static control.}
    \label{fig:circuit-static-cost}
\end{figure*}
\subsection{Inverted Pendulum}
The inverted pendulum system has state $x = [\theta, \dot \theta]$ where $\theta$ is the angle and $\dot \theta$ is the angular velocity. The control objective is to stabilize the pendulum around the upright position.

As shown in Figure \ref{fig:ip-cost} and Table \ref{table:cost}, our approach not only ensures stability but also achieves better optimality than the other methods, even NODE which solves the optimal control problem.

% In comparison, shown in Figures \ref{fig:ip-cost} and \ref{fig:pendulum-static-cost} and Table \ref{table:cost}, NODE directly solves the corresponding optimal control problem by optimizing over trajectories, which is computationally expensive and lacks stability guarantees. In contrast, our method inherently guarantees both stability and optimality, achieving results that are better than NODE.\\
We further extend our framework beyond static state feedback control to include dynamic control, which incorporates mechanisms like proportional-integral controllers. Dynamic controllers differ from static state feedback in that they consider not only the current state of the system but also past states or control actions, providing additional flexibility and improved performance for systems with memory or delayed responses. This formulation is detailed in Appendix \ref{sec:dynamic-control}, where we show that the approach studied in \cite{cui2024structured} can be viewed as a special case of our framework. As shown in Figure \ref{fig:IPd-storage}, the learned storage function and supply rate demonstrate that our method stabilizes the closed-loop system. This highlights the ability of our approach to enforce theoretical guarantees effectively.

We evaluate our method alongside baselines under the dynamic control setting. As shown in Table \ref{table:cost}, our method \text{-} both with and without cost function shaping \text{-} outperforms NCLF. This advantage likely comes from the fact that NCLF lacks guarantees of optimality, which can lead to augmented control signals and augmented states becoming large in dynamic controllers. For the static and dynamic feedback cases, our method achieves optimality comparable to or even better than NODE. However, NODE still suffers from the absence of stability guarantees.
\vspace{-5pt}
\subsection{Rod on the Cart}
We consider a flexible rod mounted on a moving cart, consisting of a cart with a metal rod fixed on top and a mass at the rod's end. Unlike traditional cart-pole systems, the rod has no joint at its base but is flexible, allowing horizontal displacement between its top and bottom positions. This case is taken from \cite{junnarkar2024synthesizing}.

The controller designed by our framework ensures fast convergence and smooth control behaviour, effectively stabilizing the flexible rod in the upright position. In terms of optimality, our method significantly outperforms NCLF, as shown in Table \ref{table:cost}. This is likely due to the fact that NCLF lacks transient performance optimization, which can lead to overshooting in the designed controller. In contrast, our dissipativity-based approach inherently incorporates optimality, particularly when cost function shaping is applied, ensuring both stability and efficient control performance.

Table \ref{table:cost} and Figure \ref{fig:circuit-static-cost} present a comprehensive cost comparison across all methods for the three systems. For each system, we conduct 30 trials with random initial conditions within the specified region of interest. The consistent low costs achieved by OURS validate the optimality of our control design.

It is worth noting that even without cost function shaping, our method inherently guarantees closed-loop stability and a certain level of optimality due to its dissipativity-based design. This is reflected by the results, where our method always outperforms NCLF in optimality. The addition of cost function shaping further enhances our framework by allowing it to adapt to a specified cost function, effectively solving particular optimal control problems.

These results emphasize the key advantage of our dissipativity-based approach: dissipativity not only ensures closed-loop stability through energy-consistent behaviour but also provides guarantees of optimality, with or without cost shaping. The cost shaping extension broadens the applicability of the framework to handle customized performance objectives, bridging stability and performance in a unified framework.

\section{Conclusion}\label{sec:conclusion}
We proposed a novel framework for designing optimal stabilizing control for nonlinear systems by learning the system's input-output dissipativity. Unlike existing learning-based control methods, our approach ensures both stability and superior optimality by design, as demonstrated on linear and nonlinear system examples.

While our reformulation reduces the sampling space dimension, learning storage and supply rate functions remains challenging due to multiple hard constraints that must be handled during training. These constraints are enforced via appropriate loss functions. Recent work on direct parameterization of robust neural networks \cite{wang2023direct} suggests a promising direction. We plan to explore parameterizing storage and supply rate functions with neural networks to enforce some constraints by design.

\section*{Acknowledgement}
The authors thank Professors Kostas Margellos, Alessandro Abate and Konstantinos Gatsis for fruitful discussions.

\bibliography{example_paper}
\bibliographystyle{ieeetr}

%%%%%%%%%%%%%%%%%%%%%%%%%%%%%%%%%%%%%%%%%%%%%%%%%%%%%%%%%%%%%%%%%%%%%%%%%%%%%%%
%%%%%%%%%%%%%%%%%%%%%%%%%%%%%%%%%%%%%%%%%%%%%%%%%%%%%%%%%%%%%%%%%%%%%%%%%%%%%%%
% APPENDIX
%%%%%%%%%%%%%%%%%%%%%%%%%%%%%%%%%%%%%%%%%%%%%%%%%%%%%%%%%%%%%%%%%%%%%%%%%%%%%%%
%%%%%%%%%%%%%%%%%%%%%%%%%%%%%%%%%%%%%%%%%%%%%%%%%%%%%%%%%%%%%%%%%%%%%%%%%%%%%%%
\appendix

\section{Dynamic Control Design}\label{sec:dynamic-control}
Dynamic control is widely used in control theory. One example is proportional-integral-derivative (PID) control, which is well-known for its capability of fast convergence (proportional), eliminating tracking errors (integral) and reducing overshoot (derivative). For linear systems, these three parts are designed linearly over states, error and the integral of states \cite{visioli2006practical}. The paradigm has been extended to nonlinear PI control for nonlinear systems, and the control can be designed with learning techniques \cite{cui2024structured}. Here, we propose a universal framework for dynamic controller design to enable more flexible and efficient operation. The diagram of the plant and the controller is shown as follows.

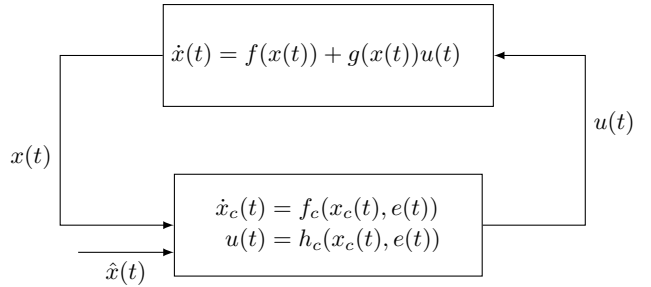
\begin{figure}[ht]
    \centering
 \begin{tikzpicture}[auto, node distance=2.5cm, >=latex, scale=0.9, every node/.style={transform shape}]

    % Nodes for the blocks
    \node [draw, rectangle, minimum width=4.5cm, minimum height=1.5cm] (system1) 
        {$\begin{aligned}
            \dot{x}(t) &= f(x(t)) + g(x(t))u(t)
        \end{aligned}$};
    
    \node [draw, rectangle, minimum width=4.5cm, minimum height=1.5cm, below of=system1] (system2) 
        {$\begin{aligned}
            \dot{x}_c(t) &= f_c(x_c(t),e(t))\\
            u(t) &= h_c(x_c(t),e(t))
        \end{aligned}$};
    
    % Arrows connecting the blocks from side centers
    \draw[->] (system1.west) -- ++(-1.5,0) -- ++ (0,-1.5) node[left] {$x(t)$} -- ++(0,-1) -- (system2.west);
    \draw[->] (system2.east) -- ++(1.5,0) -- ++ (0,1.5) node[right] {$u(t)$} -- ++(0,1) -- (system1.east);
    
    % New arrow with notation \hat{y}(t)
    \draw[->] ($(system2.west) + (-1.4, -0.4)$) -- ($(system2.west) + (0,-0.4)$) node[midway, below] {$\hat{x}(t)$};

\end{tikzpicture}
    \caption{Overall system framework: interconnected dynamical system and control system. The block on the top represents the plant, and the one on the bottom represents the controller. In the controller dynamics, $e(t):=\hat x(t)-x(t)$ represents the tracking error, where $\hat x(t)$ is a reference signal.}
    \label{fig:system-diagram-new}
\end{figure}
Compared with the static control design Framework \ref{fig:structure}, the dynamic control design Framework \ref{fig:system-diagram-new} introduces auxiliary state $x_c$ for the controller dynamics. The closed-loop system can be regarded as an interconnected system with a dynamical plant and a dynamical controller. The interconnected system can be represented by the state-space model:
\begin{equation}\label{eq:interconnected}
    \underbrace{\begin{bmatrix}
        \dot x\\
        \dot x_c
    \end{bmatrix}}_{x_e}=\underbrace{\begin{bmatrix}
        f(x)\\0
    \end{bmatrix}}_{f_e(x_e)}
    +\underbrace{\begin{bmatrix}
        0&g(x)\\I&0
    \end{bmatrix}}_{g_e(x_e)}
    \underbrace{\begin{bmatrix}
        f_c(x_c,e)\\
        h_c(x_c,e)
    \end{bmatrix}}_{u_e(x_c,e)},\quad y_e = h_e(x_c,e).
\end{equation}
Here, $h_e(x_c,e)$ is a \emph{user-defined} virtual output function, which satisfies $h_e(0,0)=0$. The control takes the form of $u_e=\pi(x_c,e)$ that stabilizes the plant. We first consider the equilibrium point $\bar x_e=[\bar x \quad \bar x_c]$ and $\bar u_e=[\bar u_c\quad \bar u]$. At the equilibrium point $\bar x_e$, we should have $\dot {\bar x}_e=0$, which implies
\begin{equation}\label{eq:equilibrium}
    \begin{split}
        f(\bar x)+g(\bar x)\bar u&=0\\
        \bar u_c&=0
    \end{split}
\end{equation}

The dissipativity condition \eqref{eq:dissi-2} can be extended for the augmented  system \eqref{eq:interconnected} with a non-zero equilibrium point $\bar x_e$:
\begin{align}\label{eq:eid}
	\nabla V_e (\bar x_e)^\top [f_e (x_e) + g_e (x_e) u_e] + \le
	(\bar y_e-y_e)^\top Q(\bar x_e) (\bar y_e-y_e) \nonumber\\ 
	+ 2(\bar y_e-y_e)^\top S(\bar x_e) (\bar u_e - {u}_e)+ (\bar u_e - {u}_e)^\top R(\bar x_e) (\bar u_e - {u}_e).
\end{align}
A similar condition has been presented in \cite{simpson2018equilibrium,madeira2022global} for static feedback control.

Similarly to the static control \eqref{eq:structured-control}, the nonlinear dynamical control $u_e$ is given by
\begin{equation}\label{eq:dyna mic-control}
    u_e = \bar u_e - R(\bar x_e)^{-1} S(\bar x_e)^\top y_e,
\end{equation}
where $\bar u_e$ is a feedforward compensator, $-R^{-1}S(\bar x_e)y_e$ is the nonlinear dynamical controller. For the special case where $y_e=x_e$, the nonlinear dynamical controller $u_e$ becomes a state feedback control
\begin{equation*}
    u_e=\bar u_e-R(\bar x_e)^{-1}S(\bar x_e)^\top x_e.
\end{equation*}
If \eqref{eq:eid} and additionally $\Delta(\bar x_e)=S(\bar x_e)R(\bar x_e)^{-1}S(\bar x_e)^\top -Q(\bar x_e)\succ 0$ hold for any $x$ and $x_c$, then stability of the augmented system \eqref{eq:interconnected} with $u_e$ in \eqref{eq:dyna mic-control} is guaranteed using Theorem \ref{th:stability}. 

Moreover, the proposed dynamical control is a generalization of nonlinear PI control. To see this, by letting 
\begin{equation}
\begin{split}
     &f_c(x_c(t),e(t))=e(t),\\
     &h_c(x_c(t),e(t))=k_I(x_c(t))+k_p(e(t))
\end{split}
\end{equation}
the controller dynamics become
\begin{equation}\label{eq:nonlinear-pi}
    \begin{split}
        x_c(t)&=\int_0^te(\tau)d\tau\\
        u(t)&=k_I(x_c(t))+k_p(e(t)),
    \end{split}
\end{equation}
where $x_c$ has a clear physical meaning, \emph{cumulative tracking error}, $k_I(x_c(t))$ is then the nonlinear integral control and $k_p(e(t))$ is the nonlinear proportional control. Neural nonlinear PI control in the form of \eqref{eq:nonlinear-pi} has been considered in \cite{cui2024structured}. Framework \ref{fig:system-diagram-new} extends it.

% \newpage
% \onecolumn
% \section{Output Feedback Passivity and Optimality}\label{sec:optimality}
\end{document}